\documentstyle[12pt]{article}
\def\lsim{\mathrel{\rlap{\raise 2.5pt \hbox{$<$}}\lower 2.5pt}}
\def\gsim{\mathrel{\rlap{\raise 2.5pt \hbox{$>$}}\lower 2.5pt}}
\newcommand{\al}{\mbox{$\alpha$}}
\newcommand{\B}{\mbox{$\beta$}}
\newcommand{\G}{\mbox{$\gamma$}}
\newcommand{\Fpi}{\mbox{$F_\pi$}}
\newcommand{\lb}[1]{\mbox{$\bar{l}_#1$}}

\begin{document}
\bibliographystyle{plain}
%\begin{titlepage}
\thispagestyle{empty}
\vspace{-3mm}
\begin{small}
\begin{flushright}
IISc-CTS/2/99\\
FZJ-IKP(TH)-1999-05
\end{flushright}
\end{small}
\begin{center}
{\bf  Pion-Pion Scattering in Chiral Perturbation and
Dispersion Relation Theories %\marginpar{\small shall we
%  delete "and Related Topics"?}
}\footnote{Work supported in part by DFG under contract no.~ME 864-15/1}
\end{center}
\vskip 2cm
\begin{small}
\begin{center}
{\bf B. Ananthanarayan} \footnote{Invited speaker at the
``Frontiers of Fundamental Physics''  Symposium, B. M. Birla
Science Centre, Hyderabad, India, December 30, 1998 - January 1, 1999}\\
Centre for Theoretical Studies, \\
Indian Institute of Science, Bangalore 560 012, India\\
%\[anant@cts.iisc.ernet.in\]\\
\vskip 1cm
{\bf P. B\"uttiker}\\
Institut f\"ur Kernphysik (Theorie), 
Forschungszentrum J\"ulich,\\
D-52425 J\"ulich, Germany\\
%\[P.Buettiker@fz-juelich.de\]]
\end{center}

\vskip 3cm

\begin{abstract}
Chiral perturbation theory, the low energy effective
theory of the strong interactions for the light pseudoscalar
degrees of freedom, is based on effective
Lagrangian techniques and is an expansion in the powers
of the external momenta and the powers of the quark masses, which
correct the soft-pion theorems.  Our primary emphasis will
be on the problem of $\pi\pi$ scattering. After briefly reviewing
these features and some results, we review some features of
$\pi-N$ scattering.
\end{abstract}
\end{small}
\newpage

\section{Introduction}
Pion-pion scattering is a problem that engaged the attention
of a generation of elementary particle physicists.  Today,
many important inputs towards a possible comprehensive understanding
of the problem requires inputs from effective Lagrangian or
chiral perturbation theory techniques, in addition to
the well-known dispersion relation techniques, suitably modified.
The purpose of this talk is to present
in a concise manner some of the results and techniques of
effective field theories that arises in the low energy
sector of the strong interactions.  The topics are the techniques and results
of modern chiral perturbation theory~\cite{ba1,zero,gl1,etc,hl1,jg1}.
Much of what will be said can also be
found in standard textbooks~\cite{sw1,dgh}.
The pion was posited to explain
the forces between two nucleons.
$\pi^\pm$
and $\pi^0$ are the lightest hadrons and are
assumed to be degenerate in mass lying in an iso-spin
triplet.  Their lightness may be understood 
by regarding them as the Goldstone bosons of 
spontaneously broken chiral symmetry of
massless QCD; the presence of non-vanishing quark
masses shifts the pion pole to $M_\pi^2=2 \hat{m} B$,
where $\hat{m}$ is the average mass of the u- and d- quarks,
$B$ is a measure of the vacuum expectation value
$< 0 | \bar{u} u | 0 > =< 0 | \bar{d} d | 0 > =-F^2_\pi B$
and $F_\pi$ is the pion decay constant~\cite{gl1}.
QCD is the theory of quark and gluon degrees of
freedom and exhibits the property of asymptotic
freedom in that at large momenta the coupling
constant becomes smaller and one may study the
theory in a perturbation series in the strong
coupling constant $\alpha_S=g^2/(4\pi)$.  The
Lagrangian is:  
\begin{equation} \nonumber
{\cal L}_{QCD}=-{1\over 4 g^2} G^a_{\mu\nu} G^{a\, \mu\nu}+\overline{q}
i\gamma^\mu D_\mu q-\overline{q}{\cal M}q.
\end{equation}
If the quark mass matrix ${\cal M}=0$ 
then the left- and right- chiral projections may
be rotated independently.   Thus associated with
say $N$ light quark flavors, we have the chiral symmetry
$SU(N)_L\times SU(N)_R$ (the $U(1)_V$ and the anomalous 
$U(1)_A$ are not considered here).
This chiral symmetry is broken spontaneously to
$SU(N)_V$, where $V=(L+R)$,
and corresponding to the $SU(N)_A$ ($A=R-L$) broken symmetry
we have $N^2-1$ (pseudoscalar) Goldstone bosons.
Although  %\marginpar{\small I don't understand the sentence "Although..."
we do not yet know how to obtain the hadronic
spectrum from the QCD Lagrangian, nor 
the mechanism by which chiral symmetry is
broken spontaneously, the QCD
Lagrangian provides the justification for the
successful results obtained from PCAC and current
algebra. The effective low energy theory of the
strong interactions at next to leading order
requires the knowledge of the underlying theory and
the analysis rests on writing down
the generating functional for the currents
of the theory which is the vacuum to vacuum
transition amplitude in the presence of external
sources.  The low energy expansion is one that
involves derivatives of the external sources
a well known example being the Euler-Heisenberg
method of analyzing QED.   
Chiral perturbation theory then is the low-energy
effective theory of the strong interactions and
involves a simultaneous expansion in the mass of
the quarks and the external momenta about the chirally
symmetric $SU(2)\times SU(2)$ limit of the massless
QCD (here we work in a world with
two quark flavors) with the the spontaneous
breakdown of this symmetry by the ground state to $SU(2)_V$, the
pions corresponding to the Goldstone bosons of
the broken $SU(2)_A$ generators.
The Goldstone theorem yields
\begin{equation}
< 0|A_\mu|\pi > =F_\pi p_\mu,
\end{equation}
and $F_\pi\approx 93$ MeV.
To leading order, $O(p^2)$, the
effective Lagrangian is that of the non-linear
sigma model.  The effective action is
\begin{equation}
Z_1=F^2\int dx {1\over 2} \nabla_\mu U^T \nabla^\mu U,
\end{equation}
where $U$ is a four component real $O(4)$
(note that $O(4)\equiv SU(2)\times SU(2)$) unit vector.
This model is not renormalizable
and the loops of the model
lead to divergences which cannot be absorbed into
the parameters of the model.  In order to absorb the
divergences, one is led to introduce higher derivative
interactions, which then allow one to extend the predictions
at leading order in the momentum or derivative to the next order.
The price is the proliferation of low energy constants (LEC) %\marginpar{\small shall we
%  call them coupling or low energy constants?}
that must be extracted from experiment or alternatively
from theoretical considerations such as the behavior
of the low energy constants %\marginpar{\small LEC?} 
in the chiral limit as well
as non-perturbative approaches such as large $N_c$ calculations.
The effective Lagrangian at $O(p^4)$~\cite{gl1} and at $O(p^6)$~\cite{fs}
have been worked out.  (When one considers the interactions
of pions with nucleons, the chiral power counting is
different since the nucleon mass does not go to zero in the chiral limit as
the pion mass does.)
For completeness we write down the effective Lagrangian at
$O(p^4)$~\cite{gl1}:
\begin{small}
\begin{eqnarray}
& \displaystyle {\cal L}_4=l_1(\nabla^\mu U^T \nabla_\mu U)^2 + l_2 
(\nabla^\mu U^T \nabla^\nu U)
(\nabla_\mu U^T \nabla_\nu U)+
l_3(\chi^T U)^2+   & \nonumber \\
& \displaystyle l_4(\nabla^\mu \chi^T \nabla_\mu U)+ 
l_5(U^T F^{\mu\nu} F_{\mu\nu} U) +
l_6 (\nabla^\mu U^T F_{\mu\nu} \nabla^\nu U)+ l_7 (\tilde{\chi}^T U)^2+ &
 \\
& \displaystyle
h_1 \chi^T \chi + h_2 {\rm tr} F_{\mu\nu} F^{\mu\nu} +h_3 \tilde{\chi}^T
\tilde{\chi}, & \nonumber
\end{eqnarray}
\end{small}
where $F_{\mu\nu}$ are covariant tensors involving the external
fields and their derivatives and the vectors $\chi$ and $\tilde{\chi}$
are proportional to the external scalar and pseudoscalar fields.
With this effective Lagrangian and with the loops generated by
the non-linear sigma model and appropriate renormalization, one
may obtain the Green's functions of QCD in
the momentum expansion.  At this order, 10 additional low energy constants
enter the effective Lagrangian.

Although the number of coupling
constants at $O(p^6)$ is very large ($>100$), those
entering the pion-pion scattering amplitude are still
limited in number.
Of course, once the coupling
constants are fixed from a certain class of experiments,
at that order,  the theory would have predictions
for all other processes at the appropriate level of
accuracy.  Furthermore, the external field technique
permits an off-shell analysis of the Green's functions
of the theory and allows one to study the quark mass
dependence of the Green's functions.

The important processes of $\pi\pi$ and $\pi N$ scattering have
been analyzed in great detail and methods have been described
in standard books~\cite{gb,mms,jlp,rgm,bm,gh2}, and will be of
interest to us in this discussion.  Note that
a good deal of the experimental information on the processes
of interest to us has been obtained via dispersion relation
analysis of phase shift information.  In fact, there is
a rich interplay between the effective Lagrangian methods
of chiral perturbation theory and dispersion relation theory
which we will describe below.   

In the following sections, we briefly 
review the status of $\pi\pi$ scattering and $\pi-N$
scattering.  
A few remarks are listed on other subjects
of interest with some references to the
literature.

\section{$\pi\pi$ Scattering}
We describe the $\pi\pi$
process~\cite{mms,gw1} in some detail, although these results
are now nearly 40 years old \footnote{Note that we assume here iso-spin to be
  a conserved quantity in strong interactions.}.
In axiomatic field theory the validity 
of dispersion relations have
been proved some time ago.  
In the case of $\pi\pi$ scattering dispersion 
relations are particularly simple.
Phase shift information has been analyzed, well 
before chiral perturbation theory or QCD were established:
an analysis that employs chiral results {\it ab initio}
is now required to confront experimental data.\\ 
Pion-pion scattering is described
in terms of a single function $A(s,t,u)$ of
the Mandelstam variables $s,t,u$ \cite{cm}.   
The process is schematically represented by
\begin{equation}
\pi^a(p_1)+\pi^b(p_2)\rightarrow \pi^c(p_3) + \pi^d(p_4)
\end{equation}
and since iso-spin is conserved by the strong
interactions, the transition matrix is given by:
\begin{equation}
A(a,b\to c,d)=A(s,t,u)\delta^{ab}\delta^{cd}+A(t,u,s)\delta^{ac}\delta^{bd}
+A(u,s,t)\delta^{ad}\delta^{bc},
\end{equation}
where the function $A(s,t,u)=A(s,u,t)\equiv A_s$ due to generalized
Bose statistics and $s=(p_1+p_2)^2$, $t=(p_1+p_3)^2$
and $u=(p_1+p_4)^2$, all momenta taken to be incoming.
$\sqrt{s}$ represents the centre of mass energy and $t$ and $u$ are related to
the cosine of the centre of mass
scattering angle via $\cos\theta=(t-u)/(s-4)$, $s+t+u=4$,
when setting the pion mass to unity.
Since the pions lie in an iso-spin
triplet, the s-channel amplitudes for definite iso-spin
can be written down:
\begin{eqnarray}
 \displaystyle T^0_s(s,t,u)& = &3 A_s + A_t + A_u  \nonumber \\
 \displaystyle T^1_s(s,t,u)& = &       A_t - A_u  \\ \label{eq:amp:iso:def}
 \displaystyle T^2_s(s,t,u)& = &       A_t + A_u  \nonumber
\end{eqnarray}
which follows from iso-spin coupling (see e.g. Ref.~\cite{burk}).
One convenient representation for dispersion relations
for the amplitudes of definite iso-spin in the t-channel
with two subtractions is:
\begin{eqnarray}
\displaystyle
T^I_t(s,t,u) & = &\mu_I(t) + \nu_I (t) (s-u)   \nonumber \\
             &  & + \displaystyle {1\over \pi} \int_4^\infty {ds' \over s'^2}
\left( {s^2\over s'-s} + (-1)^I {u^2\over s'-u}\right) \sum_{I'} C_{st}^{II'}
A^{I'}_s(s',t),
\end{eqnarray}
where $\mu_I(t), \, \nu_I(t)$ are unknown t-dependent
subtraction constants ($\mu_1=\nu_0=\nu_2=0$) (the number of subtractions is
dictated by the Froissart bound), and $A^I_s(s',t)\equiv \mbox{Im}
T^I_s(s',t)$ is the absorptive part of the s-channel amplitude.
The matrix $C_{st}$ is a so-called crossing matrix,
(embodying the fundamental property of crossing in axiomatic
field theory)
the entries of which may be written down from the general
formula resulting from iso-spin coupling in terms of
the Wigner 6-j symbol~\cite{fjd} as:
\begin{equation}
C_{st}(c,d)=(-1)^{(c+d)} (2 c + 1)\left\{
\matrix{ 1 & 1 & d \cr 1 & 1 & c \cr} \right\}.
\end{equation}

A very convenient form of dispersion relations has been
found which eliminates the unknown functions $\mu_I, \, \nu_I$
in favour of the S- wave scattering lengths $a^0_0$ and $a^2_0$ \cite{smr1}.
The scattering lengths $a^I_l$ arise in the threshold
expansion for the partial wave amplitudes ${\rm Re}f^I_l(\nu)=\nu^l
(a^I_l+b^I_l \nu + ...)$, where the latter are defined by:
\begin{eqnarray*}
  T^I_s(s,t,u)=32 \pi \sum (2l+1) f^I_l(s) P_l((t-u)/(s-4)),
\end{eqnarray*}
where $\nu=(s-4)/4$.  This form is:
\begin{eqnarray}
T^I_s(s,t)&=& \sum_{I'}
\mbox{$\frac{1}{4}$}(s\,{\bf 1}^{II'} + t\,
C_{st}^{II'} + u\, C_{su}^{II'})\,T^{I'}_s(4,0)  \\
 & &+\int_4^\infty \!ds'\,g_2^{II'}(s,t,s')\,A^{I'}_s(s',0)
+\int_4^\infty \!ds'\,g_3^{II'}(s,t,s')\,A^{I'}_s(s',t)\,
,\nonumber
\end{eqnarray}
where $g_2(s,t,s')$ and $g_3(s,t,s')$ are matrices with the crossing matrices
$C_{st},$ $ C_{su}, $ $C_{tu}$ as their building blocks \cite{smr1}.
Furthermore, $T_s(4,0)=32 \pi (a_0^0,0,a_0^2).$ Approximating the s-channel
amplitudes by the S- and P-waves and inverting the iso-spin amplitudes,
eq. (\ref{eq:amp:iso:def}), one ends up, after some algebra, with the
representation for $A(s,t,u)$ \cite{ab1}:
\begin{eqnarray}\label{Adisprecon}
\lefteqn{A(s,t,u) = \frac{32 \pi}{3}(\gamma^0_0 - \gamma^2_0)s^2 +
                16\pi \gamma^2_0 (t^2 + u^2) + 
                4\pi \al^{2}_0 (u + t)} \nonumber\\
         &   & +\frac{8\pi}{3}(\al^{0}_0 - \al^{2}_0)s + 16\pi \B^1_1 t(s-u) +
               16\pi \B^1_1 u (s-t)  \\
         &   & + 32\pi \left( {1\over 3} {s^3\over \pi}
\int_4^\infty {dx \over x^3 (x-s)}\left( {\rm Im}f^0_0(x)-
{\rm Im} f^2_0(x)\right) \right.\nonumber\\ 
&  & +\phantom{32\pi}\left. {3\over 2} (s-u) {t^2\over \pi}
\int_4^\infty {dx \over x^2 (x-t) (x-4)} {\rm Im} f^1_1(x) \right.
 \nonumber \\
& & +\phantom{32\pi}\left. {3\over 2} (s-t) {u^2\over \pi}
\int_4^\infty {dx \over x^2 (x-u) (x-4)} {\rm Im} f^1_1(x)  \right.\nonumber\\
& & +\phantom{32\pi}\left.{1\over 2} 
\left( {t^3\over \pi}\int_4^\infty {dx \over x^3 (x-t)}
{\rm Im} f^2_0(x) + {u^3\over \pi} \int_4^\infty{dx \over x^3 (x-u)}
{\rm Im} f^2_0(x) \right) \right),\nonumber
%%%%%%%%%%
\end{eqnarray}
%%%%%%%%%%
where
\begin{eqnarray}
%%%%%%%%%%
   \alpha^I_0 & = & a^I_0 -\frac{4}{\pi}\int_4^\infty\frac{dx}{x(x-4)}
                       {\rm Im}f^I_0(x) + 
                       \frac{4}{\pi}\int_4^\infty\frac{dx}{x^2}
                       {\rm Im}f^I_0(x)\quad I=0,2 \nonumber \\
   \gamma^I_0    & = & \frac{1}{\pi}\int_4^\infty\frac{dx}{x^3}
                       {\rm Im}f^I_0(x)\quad I=0,2 \label{abg} \\
   \beta^1_1    & = & \frac{3}{\pi}\int_4^\infty\frac{dx}{x^2(x-4)}
                       {\rm Im}f^1_1(x) \nonumber \\
   \alpha^1_0 & = &\gamma^1_0   = \beta^0_1=\beta^2_1=0. \nonumber 
%%%%%%%%%%
\end{eqnarray}
%%%%%%%%%%
Although the property of crossing symmetry places constraints on the
absorptive parts of the amplitude in general, the presence of two subtractions
in these dispersion relations implies that the S- and P-waves do not face any
constraints \cite{mrw}. It has been shown that the
dispersive representation for the amplitude in the approximation
that only S- and P- waves saturate the absorptive parts of
the amplitudes lends itself to a straightforward comparison
with chiral amplitudes~\cite{ab1}.   We reproduce below some of
the pertinent discussion.

Chiral perturbation theory at lowest order reproduces the current algebra
result
%%%%%%%%%%
\begin{equation}
%%%%%%%%%%
A(s,t,u)={s-1 \over  F_\pi^2}
%%%%%%%%%%
\end{equation}
%%%%%%%%%%
with only two free parameters $F_\pi$ and $m_\pi$ (note that
$m_\pi\equiv1$), leading to the prediction for $a^0_0=7/(32\pi F_\pi^2)\simeq
0.16$. Due to the non-renormalizability of the theory four
additional low energy constants $\bar{l}_{1,2,3,4}$ enter the $\pi\pi$
scattering amplitude at order $O(p^4)$, where the presence of infrared
singularities of the theory modifies the current algebra result for $A(s,t,u)$
substantially. Estimates for these quantities from disparate sources such
as D- wave scattering lengths (alternatively from $\pi\pi$
phase information directly), $SU(3)$ mass relations and
the ratio of the decay constants $F_K/F_\pi$ gives a
correction of about 25\% to the leading order prediction,
$a^0_0=0.20\pm 0.01$~\cite{gl1} (the experimental
value for this number is quoted as $0.26\pm 0.05$~\cite{nagels}).

The scattering amplitudes in chiral perturbation theory are perturbatively
unitarity; at one-loop order the loops contribute to the
scattering amplitude terms that have the correct analytic
structure corresponding to producing them from the tree
level amplitude by elastic unitarity.

%%%%%%%%%%%%%%%%%%%%%%%%%%%%%%%%%%%%%%%%%%%%%%%%%%%%%%%%%%%%%%%%
%
At next to leading order the representation for the function $A(s,t,u)$ at
$O(p^4)$ reads \cite{gl1}:
\begin{equation}
  A(s,t,u) = A^{(2)}(s,t,u) + A^{(4)}(s,t,u) +O(p^6),
\end{equation}
with
\begin{eqnarray}
A^{(2)}(s,t,u) & = & \frac{s - 1}{F^2_\pi},\nonumber\\
A^{(4)}(s,t,u) & = & \frac{1}{6 F_\pi^4}\left(3 (s^2 - 1)\bar{J}(s) 
                     \right. \nonumber \\
               &   & + \left.{[t(t-u) - 2 t + 4 u -2]\bar{J}(t) +
                     (t \leftrightarrow u)}\right ) \nonumber\\
               &   & + \frac{1}{96 \pi^2 F_\pi^4} \{2(\lb{1}-4/3)(s-2)^2 +
                     (\lb{2}-5/6)[s^2 +(t-u)^2]\nonumber\\
               &   & + 12 s (\lb{4} - 1) -3 (\lb{3} + 4 \lb{4} - 5)\}
\nonumber \\
\mbox{ and }
\bar{J}(z)  & = &  -\frac{1}{16\pi^2}\int^1_0 dx \ln [1-x(1-x)z],
\quad {\rm Im}\bar{J}(z) = {\rho(s)\over 16\pi}\Theta(z-4).  \nonumber
\end{eqnarray}
Note also that at $O(p^4)$ the imaginary parts of the partial
waves above threshold ($s>4$) computed from the amplitude above are
\begin{eqnarray}\label{chiralabs}
{\rm Im} f^0_0(s) & = & {\rho(s)\over 1024 \pi^2 \Fpi^4}(2s-1)^2 \nonumber \\
{\rm Im} f^1_1(s) & = & {\rho(s)\over 9216 \pi^2 \Fpi^4}(s-4)^2\nonumber \\
{\rm Im} f^2_0(s) & = & {\rho(s)\over 1024 \pi^2 \Fpi^4}(s-2)^2 \\
{\rm Im} f^I_l(s) & = & 0,\quad l \geq 2, \nonumber
\end{eqnarray}
(the chiral power counting enforces the property that the absorptive
parts of the D- and higher waves arise only at $O(p^8)$)
verifying the property of perturbative unitarity, viz.,
when the $O(p^2)$ predictions for the threshold parameters
$a^0_0=7/(32\pi\Fpi^2),\ a^2_0=-1/(16\pi\Fpi^2),\ b^0_0=1/(4\pi\Fpi^2),\ 
b^2_0=-1/(8\pi\Fpi^2)$ and $a^1_1=1/(24\pi\Fpi^2)$ are inserted into the
pertinent form of  the perturbative unitarity relations:
\begin{eqnarray*}
{\rm Im} f^I_0(s) & = & \rho(s)(a^I_0+b^I_0 (s-4)/4)^2,\ I=0,2 \\
{\rm Im} f^1_1(s) & = & \rho(s) (a^1_1 (s-4)/4)^2.
\end{eqnarray*}
In order to carry out the comparison between the chiral
expansion and the physical scattering data, we first recall
that up to $O(p^6)$, it is possible to decompose $A(s,t,u)$ into
a sum of three functions of single variables as follows~\cite{Stern1}:
%%%%%%%%%%
%%%%%%%%%%
\begin{eqnarray}\label{eq:a_chi}
%%%%%%%%%%
%%%%%%%%%%
& \displaystyle A(s,t,u) = 32\pi\left[
                        \frac{1}{3} W_0(s) + \frac{3}{2} (s-u) W_1(t) +
                                \frac{3}{2} (s-t) W_1(u) \right.
 & \nonumber \\
& \displaystyle
         \left.                      + \frac{1}{2} \left( W_2(t) + W_2(u) - 
                                \frac{2}{3} W_2(s)\right)\right]. &
%%%%%%%%%%
%%%%%%%%%%
\end{eqnarray}
%%%%%%%%%%
%%%%%%%%%%
One convenient decomposition of the chiral one-loop amplitude is:
\begin{eqnarray}
W_0(s) & = & {3 \over 32 \pi }
\left[ {s-1 \over F_\pi^2} +
{2 \over 3 F_\pi^4} (s-1/2)^2 \bar{J}(s)  \right. \nonumber \\
& &\left. + {1\over 96 \pi^2 F_\pi^4}
(  2 (\lb{1}-4/3) (s-2)^2 + 4/3 (\lb{2}-5/6)(s-2)^2 \right. \nonumber \\
& &\left. + 12 s (\lb{4}-1)-3(\lb{3}+4 \lb{4}-5))\phantom{{s-1 \over F_\pi^2}}
\hspace{-8mm}\right],\\  
W_1(s) & = & {1\over 576 \pi F_\pi^4}(s-4)\bar{J}(s),  \\
W_2(s) & = & {1\over 16 \pi}\left[
{1\over 4 F_\pi^4}(s-2)^2 \bar{J}(s)+{1\over 48 \pi^2 F_\pi^4}
(\lb{2}-5/6)(s-2)^2\right], 
\end{eqnarray}
where we note that this decomposition is not unique, with
ambiguities in the real part only.
We observe that the imaginary parts of these functions verify
the relation:
\begin{eqnarray}
{\rm Im} W_I(x)& = &{\rm Im} f^I_0(x),\quad I=0,2 \label{cuts} \\
{\rm Im} W_1(x)& = &{\rm Im} f^1_1(x)/(x-4), \nonumber
\end{eqnarray}
which may be used to demonstrate the following dispersion relations:
\begin{eqnarray}
W_0(s) & = & {-1+72\lb{1}+48 \lb{2}-27\lb{3}-108\lb{4}-864\pi^2\Fpi^2
             \over 9216 \pi^3 \Fpi^4}  \nonumber \\
       &   & + {59-144\lb{1}-96\lb{2}+216\lb{4}+1728\pi^2\Fpi^2\over 18432
             \pi^3\Fpi^4}s   \\
       &   & + {-797+360\lb{1}+240\lb{2}\over 184320\pi^3\Fpi^4} s^2 +
             {s^3\over \pi}\int_4^\infty {dx\over x^3 (x-s)} {\rm Im}
             f^0_0(x), \nonumber \\
W_1(s) & = & {-s \over 13824 \pi^3 \Fpi^4}+{s^2\over \pi}
              \int_4^\infty {dx \over x^2 (x-4) (x-s)}{\rm Im} f^1_1(x), \\
W_2(s) & = & {6\lb{2}-5 \over 1152 \pi^3 \Fpi^4} + 
             {23-24\lb{2} \over 4608 \pi^3 \Fpi^4}s + {60 \lb{2} -77 \over
              46080 \pi^3 \Fpi^4}s^2  \\
       &   & + {s^3\over\pi} \int_4^\infty
             {dx \over x^3 (x-s)}{\rm Im} f^2_0(x). \nonumber
\end{eqnarray}
We now reconstruct $A(s,t,u)$
from this dispersive representation for the $W$'s to obtain:
\begin{small}
\begin{eqnarray}\label{Achirecon}
A(s,t,u) & = & {s-1\over \Fpi^2}+{-540+480\lb{1}+960\lb{2}- 180\lb{3}-
                720\lb{4} \over 5760 \pi^2 \Fpi^4}  \nonumber \\
         &   & -{110+480 \lb{1}-720 \lb{4}\over 5760 \pi^2 \Fpi^4}s -
               {163-120\lb{1}\over 5760 \pi^2 \Fpi^4}s^2  \nonumber \\
         &   & +{460-480\lb{2}\over 5760 \pi^2 \Fpi^4}(t+u) -{20u(s-t)
                +20t(s-u)\over 5760 \pi^2 \Fpi^4}  \nonumber \\
         &   & -{154-120\lb{2}\over 5760 \pi^2 \Fpi^4}(t^2+u^2)  \\
         &   & +32\pi \left( {1\over 3} {s^3\over \pi}\int_4^\infty
               {dx \over x^3 (x-s)}\left( {\rm Im}f^0_0(x)-
               {\rm Im} f^2_0(x)\right)  \right.\nonumber\\
         &   & \left. +{3\over 2} (s-u) {t^2\over \pi}
               \int_4^\infty {dx \over x^2 (x-t) (x-4)}
               {\rm Im} f^1_1(x) \right. \nonumber \\
         &   & \left. +{3\over 2} (s-t) {u^2\over \pi}\int_4^\infty
               {dx \over x^2 (x-u) (x-4)} {\rm Im} f^1_1(x) \right.\nonumber\\
         &   & \left. +{1\over 2} \left( {t^3\over \pi}\int_4^\infty
               {dx \over x^3 (x-t)}{\rm Im} f^2_0(x) + {u^3\over \pi}
               \int_4^\infty{dx \over x^3 (x-u)}
               {\rm Im} f^2_0(x) \right ) \right).\nonumber
\end{eqnarray}
\end{small}
This is seen to be the sum of a polynomial of second degree in
$s$, $t$ and $u$ and a dispersive piece.  The 
problem associated with the non-uniqueness of
the real part of the decomposition into the $W$'s is eliminated by setting
$u=4-s-t$ upon which we obtain a second degree polynomial in
$s$ and $t$:
%%%%%%%%%%
\begin{small}
\begin{eqnarray}\label{eq:chiral:poly}
%%%%%%%%%%
\lefteqn{P = \left( \frac{29}{120\pi^2\Fpi^4} - 
             \frac{\lb{2}}{6\pi^2\Fpi^4}\right)
             \left(t - {t^2\over 4}  - {s t\over 4}\right)}\nonumber\\
  &   & +\left( -\frac{33}{640\pi^2\Fpi^4} + \frac{\lb{1}}{48\pi^2\Fpi^4} + 
         \frac{\lb{2}}{48\pi^2\Fpi^4} \right) s^2 \nonumber \\
  &   & +\left(\frac{1}{\Fpi^2} + \frac{97}{960\pi^2\Fpi^4}-\frac{\lb{1}}
         {12\pi^2\Fpi^4}-\frac{\lb{2}}{12\pi^2\Fpi^4}
         +\frac{\lb{4}}{8\pi^2\Fpi^4} \right) s \\
  &   &  +\left(-\frac{1}{\Fpi^2} - \frac{97}{480\pi^2\Fpi^4}
         +\frac{\lb{1}}{12\pi^2\Fpi^4}+\frac{\lb{2}}{6\pi^2\Fpi^4}
         -\frac{\lb{3}}{32\pi^2\Fpi^4}-\frac{\lb{4}}{8\pi^2\Fpi^4} \right).
         \nonumber
%%%%%%%%%%
\end{eqnarray}
\end{small}
%%%%%%%%%%
We now wish to employ the knowledge that the low energy representation
is fully determined by the two subtraction constants $a^0_0$ and $a^2_0$
and in terms of the three functions of single variables $W_I$ with
right hand cuts only and verifying eq.(\ref{cuts}).  Indeed,
a Roy equation fit allows us to obtain a representation for
the S- and P- wave absorptive parts, (with some effects of higher
angular momentum states absorbed
into the driving terms).

We are now in a position to compare the two representations
for $A(s,t,u)$ namely the chiral representation eq.(\ref{Achirecon})
and the axiomatic representation eq.(\ref{Adisprecon}).
These are formally equivalent,
with the dispersive integrals in the former
described by chiral absorptive parts whereas in the latter by the physical S-
and P-wave absorptive parts.  For the chiral expansion to reproduce low energy
physics accurately we now require the effective subtraction constants to
match.  Once more setting $u=4-s-t$ yields the polynomial piece of
the representation eq.(\ref{Adisprecon}):
%%%%%%%%%%
\begin{eqnarray}\label{eq:disp:poly}
%%%%%%%%%%
P & = & -128\pi(\B^1_1 + \G^2_0)( t - {t^2\over 4} - {s t\over 4})+
        \frac{16\pi}{3}(2 \G^0_0 + \G^2_0 - 3 \B^1_1) s^2 
        \nonumber\\
  &   &  + 8\pi ({\al^0_0\over 3} - {5\over 6}\al^2_0 + 8\B^1_1 - 16\G^2_0)s
+ 16\pi (\al^2_0 + 16 \G^2_0).
%%%%%%%%%%
\end{eqnarray}
%%%%%%%%%%
A straightforward comparison of eq. (\ref{eq:chiral:poly}) and
eq. (\ref{eq:disp:poly}) 
yields
explicit expressions for \lb{1}, \lb{2}, \lb{3} and \lb{4}.
In particular we have for \lb{1} and \lb{2}:
\begin{eqnarray}
\lb{1} & = & 24\pi^2\Fpi^4 ({41\over 960\pi^2\Fpi^4}-{64\pi
\over 3}(\gamma^2_0-\gamma^0_0+3\beta^1_1)),\label{eq:rel1} \\
\lb{2} & = & 24\pi^2\Fpi^4 ({29\over 480 \pi^2 \Fpi^4}+32\pi(
\beta^1_1+\gamma^2_0)).\label{eq:rel2} 
\end{eqnarray}
For the numerical values we find for \lb{1} and \lb{2} \cite{ab1}:
%%%%%%%%%%
\begin{eqnarray*}
%%%%%%%%%%
  \bar{l}_1 & = & -1.7 \pm 0.15\\
  \bar{l}_2 & = & 5.0.
%%%%%%%%%%
\end{eqnarray*}
%%%%%%%%%%
These have an interesting dependence on the actual physical
phase shifts:  one observes that in eq.(\ref{eq:rel1}), 
the presence of $\gamma^0_0$.  As a result we can anticipate
\lb{1} to be influenced by the input for $a^0_0$.
In contrast, \lb{2}
has no dependence on $\gamma^0_0$ and
depends almost totally on the P- wave contribution via $\beta^1_1$,
as a result of the weakness of the $I=2$ channel which
renders $\gamma^2_0$ negligible in comparison with $\beta^1_1$
(and $\gamma^0_0$).
Since the P- wave happens to be the best determined experimental
quantity, even the Roy equation fits to it are not strongly
influenced by the input value of $a^0_0$.  Thus we expect
a determination of \lb{2} in this manner to be very stable.
%%%%%%%%%%%%%%%%%%%%%%%%%%%%%%%%%%%%%%%%%%%%%%%%%%%%%%%%%%%%%%%%%%%

The higher partial
waves have to be treated with care.  The problem must be accounted
for when we saturate fixed-t dispersion relations with absorptive
parts which are modeled theoretically (perhaps in terms of
resonance propagators, Pomeron exchange, say the Veneziano model, etc.)
A recent analysis of these inputs has been reported~\cite{ab2}.
Alternatively, one may write down dispersion relations in terms
of homogeneous variables~\cite{mrw}
which manifestly enforce crossing symmetry;
however there might be a dependence on parameters which parametrize
the curves in the plane of the homogeneous variables on which
the dispersion relations are written down.   Recently
this method has been exploited to extract the contributions
of some resonances~\cite{ba2}.  
A discussion concerning the evaluation of higher threshold
parameters of pion-pion scattering has been presented in
some detail~\cite{ab3}.

In particular, today the chiral amplitudes beyond one loop
have been computed~\cite{sternetal} and to two loops~\cite{bcegs}
(also see Ref.~\cite{gw2}).
The work of~\cite{bcegs} in standard chiral perturbation theory
affords an accurate prediction for the parameter $a^0_0$ which
is a soft quantity from the point of view of dispersion relations
and work is in progress to this end.  For some of the latest
information on the experimental as well as theoretical aspects
of the subject, see Ref.~\cite{pipiwg}.

\section{Further processes}
$\pi N$ scattering is a further well known process represented by
%%%%%%%%%%
\begin{equation}
%%%%%%%%%%
\pi^a(q_1)+N(p_1)\to \pi^b(q_2) + N(p_2).
%%%%%%%%%%
\end{equation}
%%%%%%%%%%
The pion-nucleon system has been studied in detail with the methods of current
algebra, PCAC, and early-days phenomenological Lagrangians. An excellent
introduction and review of these topics may be found in Chapter 19 of 
Ref.~\cite{sw1}.  For instance, we have the lowest order
predictions for the iso-spin 3/2 and 1/2 scattering
lengths of -0.075 and 0.15 which compare favorably with the
data~\cite{nagels}.

The $\pi N$ amplitude may be written in terms of the four invariant amplitudes
$A^\pm, B^\pm$:
%%%%%%%%%%
\begin{eqnarray}
%%%%%%%%%%
& \displaystyle T_{ab}=T^+ \delta_{ab}-T^- i \epsilon_{abc} \tau_c
& \nonumber \\
& \displaystyle T^\pm =\overline{u}(p_2) \left[ A^\pm(\nu,t) +\frac{1}{2}\gamma_\mu
  (q^\mu_1+q^\mu_2) B^\pm(\nu,t) \right] u(p_1) &
%%%%%%%%%%
\end{eqnarray}
%%%%%%%%%%
with $s=(p_1+q_1)^2, \, t=(q_1-q_2)^2, \, u=(p_1-q_2)^2, \, \nu=(s-u)/(4m_N)$.

In chiral perturbation theory, this process has been first considered
in a relativistic framework in~\cite{gss}. In this work, the authors have
extended the analysis of the Green's functions of QCD with an external nucleon.
Furthermore %\marginpar{\small Which environment?} 
the $\pi N$ environment is an
important one for the tests of chiral predictions~\cite{gh,ge}. However, the
fact that the nucleon has a non-vanishing mass in the chiral limit causes some
difficulties: the presence of the chiral-limit nucleon mass in
the lowest order $\pi-N$ Lagrangian makes it necessary to renormalize the the
chiral-limit nucleon mass. This destroys the correspondence between the number
of loops and the powers of the external momenta and makes the bookkeeping of
the chiral expansion somewhat difficult. 
These difficulties are circumvented in heavy baryon chiral perturbation theory
(HBChPT) \cite{jm}, where the nucleon mass is taken to be heavy compared to the
external momenta. This way, the nucleon mass does not show up in the lowest
order Lagrangian restoring the chiral counting scheme at the price of loosing
the manifest Lorentz invariance\footnote{Note that in a recent work it is
  shown that both chiral counting scheme and Lorentz invariance can be
  preserved \cite{bl}.} .

Both methods have in common that because of the non-renormalizability
additional low energy constants have to be introduced at each order of the
chiral expansion. Again, they can not be determined by the theory alone.
Only a comparison with experimental data allows one to fix the low energy
constants.
As an example we consider HBChPT. There one usually expresses the amplitudes
in terms of the isoscalar/isovector non-spin-flip amplitudes $g^\pm$ and the
isoscalar/isovector spin-flip amplitudes $h^\pm$:
%%%%%%%%%%
\begin{eqnarray}\label{eq:pinampl1}
%%%%%%%%%%
   A^\pm & = & C_1 g^\pm + C_2 h^\pm\\ \label{eq:pinampl2}
   B^\pm & = & C_3 g^\pm + C_4 h^\pm
%%%%%%%%%%
\end{eqnarray}
%%%%%%%%%%
with suitably chosen coefficients $C_i$.
To order $O(q^3)$ the amplitude $g^\pm$, e.g., reads \cite{FMS}:
%%%%%%%%%%
\begin{small}
\begin{eqnarray}
%%%%%%%%%%
   g^+_{tree}(\omega,t) & = & -\frac{g_A^2}{F_\pi^2}\frac{1}{16m\omega^2} \biggl[ 4M^4 +
                    t^2 + 4 \omega^2 t - 4M^2 t \biggr]\nonumber \\
              &   & + \frac{g_A^2}{F_\pi^2} \frac{1}{32m^2\omega^3}
                    \biggl[16\omega^2 M^4 + 5\omega^2 t^2 - 16M^6 -8M^2t^2 +
                    4\omega^4 t + 20 M^4 t\nonumber \\
              &   & \qquad\qquad\qquad + t^3 -
                    20 \omega^2 M^2 t \biggr]~,\\
   g^+_{ct}(\omega,t)   & = & \frac{1}{F_\pi^2}\Bigl[ -4c_1 M^2 + 2c_2 \omega^2
                    + c_3(2M^2-t) \Bigr] + \frac{c_2 \, \omega}{m F_\pi^2} 
                    \Bigl[ 4 \omega^2 - 4M^2 +t \Bigr]~,\nonumber \\
   g^+_{\rm loop}(\omega,t) &=& i\,\frac{\omega^2}{8 \pi F_\pi^4}\sqrt{\omega^2
                      -M^2} +  \nonumber \\
               & &       \frac{g_A^2}{F_\pi^4}\frac{1}{32 \pi}
                      \Bigl( M^2 -2t \Bigr)\Bigl(M+ \frac{2M^2-t}{2\sqrt{-t}}
                      \arctan \frac{\sqrt{-t}}{2M}\Bigr)\nonumber
%%%%%%%%%%
\end{eqnarray}
%%%%%%%%%%
\end{small}
where the LEC's $c_1, c_2, c_3$ show up in the counter term contribution
$g_{ct}^+$. $\omega$ is the cm energy of the pion and $M$ and $m$ are the pion
and the nucleon mass, respectively.
The above chiral calculation is most reliable inside the Mandelstam triangle,
which is a unphysical part of the Mandelstam plane. The only way to
extrapolate the experimental information to this region is to use dispersion
relations. By using crossing symmetry, unitarity and analyticity the real part
of $A^\pm$ and $B^\pm$ may be written as
%%%%%%%%%%
\begin{eqnarray}
%%%%%%%%%%
   \mbox{Re }A^\pm(s,t) & = & \frac{1}{\pi} P\int_{s_{th.}}^\infty ds'
   \mbox{Im }A^\pm(s',t)
   \left\{\frac{1}{s'-s}\pm\frac{1}{s'-u}\right\}\\
   \mbox{Re }B^\pm(s,t) & = & \frac{g_{\pi N}^2}{m^2-s}\mp\frac{g_{\pi N}^2}
   {m^2-u} \nonumber \\ 
   &  &+ \frac{1}{\pi} P \int_{s_{th.}}^\infty ds' \mbox{Im
   }B^\pm(s',t)\left\{\frac{1}{s'-s}\mp\frac{1}{s'-u}\right\},
%%%%%%%%%%
\end{eqnarray}
%%%%%%%%%%%
%
where $g_{\pi N}$ is the pion-nucleon coupling constant and
$s_{th.}=(m+M)^2$. The inclusion of all the aspects of dispersion relations
can be easily found in the literature~\cite{gh2}.
The advantage of the above relations is obvious: as the range of
integration is restricted to the physical domain of $s$, the
integrands of the dispersion integrals can be calculated from the
available experimental data. On the other hand, the dispersion
relations are valid for any s and t, i.e. by using the data in the
{\it physical} domain one is able to extrapolate
the amplitudes $A^\pm (s,t)$ and $B^\pm (s,t)$ to the {\it unphysical}
region.
By inverting eq. (\ref{eq:pinampl1},\ref{eq:pinampl2})  the results of a dispersive analysis may then be
compared with chiral expressions for the amplitudes $g^\pm$ and $h^\pm$
\cite{buemei}, yielding numerical results for the LEC's. 

The methods of chiral perturbation theory are also of great
use for strong, semi-leptonic, electromagnetic, etc. decays.
One example where all the methods of effective
Lagrangian as well as those of dispersion relations are
of utility are in the decay $\eta\to 3\pi$~\cite{etadecay}.

\section{Afterword}%\marginpar{\small Should be edited before submitting.}
In this talk we have described in some detail the effective
field theories describing a sector of the standard model
where conventional perturbation theory fails.  
Questions still remain in the sense of a rigorous
equivalence of solving the underlying field theory
and its effective field theory.  In a somewhat
indirect context these questions have recently been raised~\cite{ba3}.
The ideas and techniques used are manifold and present a great challenge
to the student and avid reader which promises sharp quantitative
and qualitative understanding of otherwise intractable phenomena.

\bigskip

\noindent{\bf Acknowledgement:}  It is a pleasure to thank the B. M. Birla
Science Centre, Hyderabad, India and its director Dr. B. G. Sidharth 
for giving us an opportunity to present a comprehensive review of
the developments in this field.

\newpage
\begin{small}

\end{small}

\end{document}